\documentclass[runningheads]{llncs}
\usepackage{graphicx}
\usepackage{todonotes}
\usepackage{amssymb}
\usepackage{amsmath}
\usepackage{hyperref}

\begin{document}
\title{Patient-specific Conditional Joint Models of Shape, Image Features and Clinical Indicators}%\thanks{Supported by organization x.}}
%/ Patient specific statistical shape models }
%
\titlerunning{Patient-specific Conditional Joint Models}
% If the paper title is too long for the running head, you can set
% an abbreviated paper title here
%
\author{
Bernhard Egger \inst{1} \and
Markus D. Schirmer\inst{1,2,3} \and
Florian Dubost\inst{2,4}\and
\\ 
Marco J. Nardin\inst{2} \and
Natalia S. Rost\inst{2} \and
Polina Golland\inst{1}
}
% index{Egger, Bernhard}
% index{Schirmer, Markus D.}
% index{Dubost, Florian}
% index{Nardi, Marco J.}
% index{Rost, Natalia S.}
% index{Golland, Polina}
%
\authorrunning{Egger et al.}
% First names are abbreviated in the running head.
% If there are more than two authors, 'et al.' is used.
%
%\institute{Princeton University, Princeton NJ 08544, USA \and
%Springer Heidelberg, Tiergartenstr. 17, 69121 Heidelberg, Germany
%\email{lncs@springer.com}\\
%\url{http://www.springer.com/gp/computer-science/lncs} \and
%ABC Institute, Rupert-Karls-University Heidelberg, Heidelberg, Germany\\
%\email{\{abc,lncs\}@uni-heidelberg.de}}
%
\institute{Computer Science and Artificial Intelligence Lab (CSAIL), MIT, USA \and
Massachusetts General Hospital (MGH), Harvard Medical School, USA \and
German Centre for Neurodegenerative Diseases (DZNE), Germany \and
Erasmus MC - University Medical Center Rotterdam, Netherlands
}

%Changelog

% Authors, Affiliations, Keywords and Acknoledgments added
% Images are smaller to not violate the template but stay within 8 pages with affiliation and acknoledgments
% References are cleaned
% Reviewer 1: suggested some clarification and corrections which where added to the paper
% Reviewer 3: Figure 2 was broken - alignment issues fixed
% Reviewer 3: Fixed wording for validation set
% Reviewer 5: proper description of cohort was removed for blind review and is now added
% Reviewer 5: number of random draws was specified
% Reviewer 6: we improved the abstract based on those comments

% We incorporated most comments by the reviewers. However the strict page limit does not allow to include all of them. We are however preparing a extended journal version and those comments will be adressed there.

% Added affiliations to supplementary material

\maketitle              % typeset the header of the contribution
\begin{abstract} %150-200 words

We propose and demonstrate a joint model of anatomical shapes, image features and clinical indicators for statistical shape modeling and medical image analysis. The key idea is to employ a copula model to separate the joint dependency structure from the marginal distributions of variables of interest. This separation provides flexibility on the assumptions made during the modeling process. The proposed method can handle binary, discrete, ordinal and continuous variables. We demonstrate a simple and efficient way to include binary, discrete and ordinal variables into the modeling. We build Bayesian conditional models based on observed partial clinical indicators, features or shape based on Gaussian processes capturing the dependency structure. We apply the proposed method on a stroke dataset to jointly model the shape of the lateral ventricles, the spatial distribution of the white matter hyperintensity associated with periventricular white matter disease, and clinical indicators. The proposed method yields interpretable joint models for data exploration and patient-specific statistical shape models for medical image analysis.

\end{abstract}

\section{Introduction}
In medical imaging almost every image is at least weakly labeled with clinical indicators. A minimal set of such labels usually includes age, sex, demographics, and diagnostic or prognostic clinical values such as blood pressure, history of smoking or outcome scores. The labels are of interest for clinical decisions or clinical research, however, they are rarely included in the analysis of shape or image features. We propose to construct joint statistical shape and attribute models to better understand and visualize the interplay between shape deformations and the variations of clinical indicators. The proposed model promises to reveal associations between anatomical shapes, image features, and clinical phenotypes, to help understand disease mechanisms and risk factors, and to improve predictions and patient stratification in clinical trials.

We demonstrate our approach in the context of a clinical cohort of ischemic stroke patients, where we investigate the relationships between the patterns of expansion of the lateral ventricles, the spatial patterns of white matter hyperintensity (WMH) indicative of the white matter disease, and patient data that include stroke outcome scores. Our model is built from segmentations of ventricles and white matter hyperintensities in T2-FLAIR  magnetic resonance images.

We integrate copula \cite{sklar1959fonctions} and Gaussian processes \cite{rasmussen2003gaussian} to jointly analyze shape and other variables. Copula methods provide a decomposition of every continuous and multivariate distribution into its marginal distributions and a copula that captures the dependency structure \cite{sklar1959fonctions}. The separate modeling of the marginal distributions and of the copula offers better flexibility than the classical statistical shape models. The marginal distributions can be modeled as Gaussian, or other parametric and non-parametric distributions. The choice of the model for each marginal distribution is separate from all others, e.g., a Gaussian distribution for the shape components and the empirical distributions for the clinical indicators.
In this work, we use a Gaussian copula, which means we transform the data into a space where all marginal distributions are Gaussian. In this latent space, we then apply standard shape modeling techniques, such as the principal component analysis (PCA) or the Gaussian processes, to model the dependency structure between 3D coordinates of the shape and other indicators. The transformation to the latent space is invertible and our shape model remains generative. We can, therefore, construct the conditional distribution given clinical indicators using Gaussian process regression. Here we focus on the setting where some clinical indicators are given and we capture the conditional distribution of shapes, i.e., a patient-specific statistical shape model.

The main contributions of this paper are: (i) a novel approach for generating conditional shape models based on clinical indicators or  image features; (ii) copula-based shape models that include binary, discrete, ordinal and continuous variables; (iii) novel associations between ventricle growth, WMH patterns and clinical indicators in the ischemic stroke patient cohort.

\textbf{Related Work} Although many clinical indicators are collected and available with medical images, the idea of combining statistical shape models with clinical indicators is rarely explored. The work of Blanc \textit{et al.} \cite{blanc2012statistical} includes patient attributes in femur modeling. The method assumes the data is Gaussian distributed and cannot handle binary, discrete, or ordinal variables. Pereanez \textit{et al.} \cite{pereanez2015patient} proposed constraining shape models based on clinical indicators. They demonstrated that such constrained models are beneficial for model-based image segmentation. This approach is limited to a model prediction given a complete set of clinical indicators and handles ordinal variables as a set of binary attributes. Our method combines these ideas to employ the copula and uses the conditional distribution to capture the remaining variability given partial data. Han and Liu \cite{han2012semiparametric} proposed a variant of PCA called Copula Component Analysis that is inherently scale-invariant, more robust to outliers and handles arbitrary discrete ordinal marginal distributions. Hoff \cite{hoff2007extending} presented the idea of an extended rank likelihood which can also account for non-continuous variables. In computer vision, Egger \textit{et al.} \cite{egger17copula,egger2016copula} used copulas for shape and appearance modeling to account for the non-Gaussian appearance of human faces, however, they did not handle discrete or ordinal variables. 

Conditional shape models without clinical indicators are well studied. A Bayesian approach was proposed by Albrecht \textit{et al.} \cite{albrecht2013posterior} to model the remaining variability of the femur given partial information.  They showed that (probabilistic) PCA and Gaussian process regression with a statistical kernel are equivalent for such modeling. We build upon this model and include clinical indicators to construct patient-specific conditional shape models based on image features and/or clinical indicators. In our experiments, we visualize the conditional models and demonstrate superior performance in a reconstruction task when including all data in the statistical model.

\section{Methods}
We propose a joint model for anatomical shape, image features and clinical indicators based on the copula \cite{sklar1959fonctions} and combine it with the idea of conditional shape models based on Gaussian processes \cite{albrecht2013posterior}. To enable the use of not only continuous, but also binary, discrete and ordinal variables, we use a sampling strategy similar to \cite{hoff2007extending}.

We define an instance of our model to be a combination of 3D shape coordinates $x_k^{(1)},x_k^{(2)},x_k^{(3)}$ and image-based feature $f_k$ associated with surface point $k$. All $N$ surface points are in correspondence across training data, achieved via shape registration to an atlas in our experiments. In addition each instance includes $K$ binary, discrete, ordinal or continuous variables $\{a_1 \ldots a_K\}$.
Let $\allowbreak y = [x_1^{(1)},x_1^{(2)},x_1^{(3)},\ldots, \allowbreak x_N^{(1)},x_N^{(2)},x_N^{(3)}, \allowbreak f_1,\ldots,\allowbreak f_N, a_1,\ldots,a_K]^T$ be the vector representation of the instance, $y \in \mathbb{R}^{d}$ with $d = 4N + K$. 
The training set of $M$ shapes with image features and clinical indicators forms the data matrix $Y \in \mathbb{R}^{d \times M}$. Let random variable $y_i$ represent component $i$ of the instance and $W_i = p(y_i)$ represents its marginal distribution.  

We use a Gaussian copula to capture the dependency pattern in combination with marginal distributions $W_i$ \cite{genest1995semiparametric,tsukahara2005semiparametric}. According to Sklar's theorem \cite{sklar1959fonctions} the copula $C$ provides the dependency structure in the joint cumulative distribution function (CDF) $F$ such that 
%remove sclars theorem at all?
\begin{equation}\label{eq:copula}
    F(y_1, \cdots, y_{n}) = C \left(W_1 , \dots, W_{n} \right) =  \Phi_R( \Phi^{-1}(W_1),\ldots, \Phi^{-1}(W_n))
\end{equation}
Where $\Phi^{-1}(\cdot)$ is the inverse standard normal CDF and $\Phi_R(\cdot)$ is the joint CDF of a zero-mean multivariate Gaussian distribution with covariance matrix $R$. The multivariate latent representation $ \hat{y}_i = \Phi^{-1} \left( W_i \right), i = 1, \ldots, d$ distributed according to the standard normal distribution. 

\textbf{Marginal Distributions} To learn the model above from training data we first estimate the marginal distributions with parametric or nonparametric methods. For properties that cannot be naturally captured by a Gaussian distribution (e.g., image features, sex or outcome scores), we employ the empirical marginal distribution for all the components as proposed in \cite{han2012semiparametric}. If only limited data is available, various assumptions on the marginal distribution must be made, separately for each component. In statistical shape modeling, a Gaussian marginal distribution is usually assumed. To transform our data into the latent space, we first transform the empirical marginal distribution into a uniform distribution by replacing every entry of the data by the CDF of the empirical distribution. Second, we use the inverse CDF of a standard normal distribution to map to the latent space \cite{egger17copula}. 

While the original copula framework is limited to continuous distributions, multiple solutions have been proposed to overcome this limitation. The first step, the transformation to a uniform distribution, relies on unique ranks (values of CDF). For binary, discrete and ordinal variables this sorting is not unique. In our experiments, we observed that the latent representation is stable when permuting elements with the same values for ranking, in the setting where we have many more variables than samples ($d \gg M$). To construct a consistent dependency structure we generate multiple (50) random rankings for non-unique values and average the resulting estimates of the model parameters. Once the data is transformed to its latent representation we estimate the covariance matrix $R$ and represent it by its $M-1$ eigenvectors (i.e. principal components).

\textbf{Conditional modeling} The Gaussian copula represents the dependency structure of our data $y$ based on the latent representation $\hat{y}$. We use Gaussian processes to capture correlations in the latent space \cite{rasmussen2003gaussian}. We define function $s_j : \Omega \rightarrow \mathbb{R}$ on a finite domain $\Omega$ such that $s_j(\hat{y}_i) \in \mathbb{R}$ refers to the $i$th element of the $j$th training example in our latent space, $i = 1, \ldots, d$, $j=1,\ldots,M$. The Gaussian process is fully defined by the mean and covariance functions estimated from the training data:
\begin{equation}
\label{eq:mean}
\begin{aligned}
\mu (\hat{y}) & = \frac{1}{M} \sum_{j=1}^{M} s_j(\hat{y})\\
\Sigma (\hat{y}, \hat{y}') & = \sum_{j=1}^{M} \Big(s_j(\hat{y})-\mu (\hat{y})\Big)\Big(s_j(\hat{y}')-\mu (\hat{y'})\Big)^T.
\end{aligned}
\end{equation}

Given a subset $z$ of the vector $y$ as an observation, we construct component-wise its latent representation $\hat{z}$. The prediction of the missing parts reduces to a standard Gaussian process regression problem. We assume observations are corrupted by uncorrelated Gaussian noise $\epsilon = \mathcal{N}(0, \sigma^2)$ and obtain the prediction for the distribution of the complete vector $\hat{y}$ from the observed vector $\hat{z}$. The conditional distribution $p(\hat{y}|\hat{z})$ is again a Gaussian process with mean:

\begin{equation}\label{eq:meanC}
\mu_c (\hat{y}|\hat{z}) = \mu (\hat{y}) + \Sigma(\hat{y},\hat{z})^T  \Big(\Sigma(\hat{z},\hat{z})+\sigma^2\mathcal{I}\Big)^{-1}  (\hat{z} - \mu_{\hat{z}})
\end{equation}
and covariance
\begin{equation}\label{eq:covarianceC}
\Sigma_c (\hat{y}, \hat{y}'|\hat{z}) = \Sigma(\hat{y}, \hat{y}') - \Sigma(\hat{y},\hat{z})^T  \Big(\Sigma(\hat{z},\hat{z})+\sigma^2\mathcal{I}\Big)^{-1}  \Sigma(\hat{z},\hat{y}')
\end{equation}
where $\mathcal{I}$ is the identity matrix.

The parameter $\sigma$ has an impact on the accuracy of the prediction and varies between different observations. The different modalities in our model have a different amount of uncertainty. For shape observations, $\sigma$ is an estimate of the segmentation accuracy and the registration quality, for clinical indicators, it is a measurement of how accurate they are measured. We estimated the optimal parameters for observation uncertainty $\sigma$ for each set of observations via cross-validation.

\section{Experiments}
\begin{figure}[t!]
\begin{center}
\includegraphics[width=\textwidth]{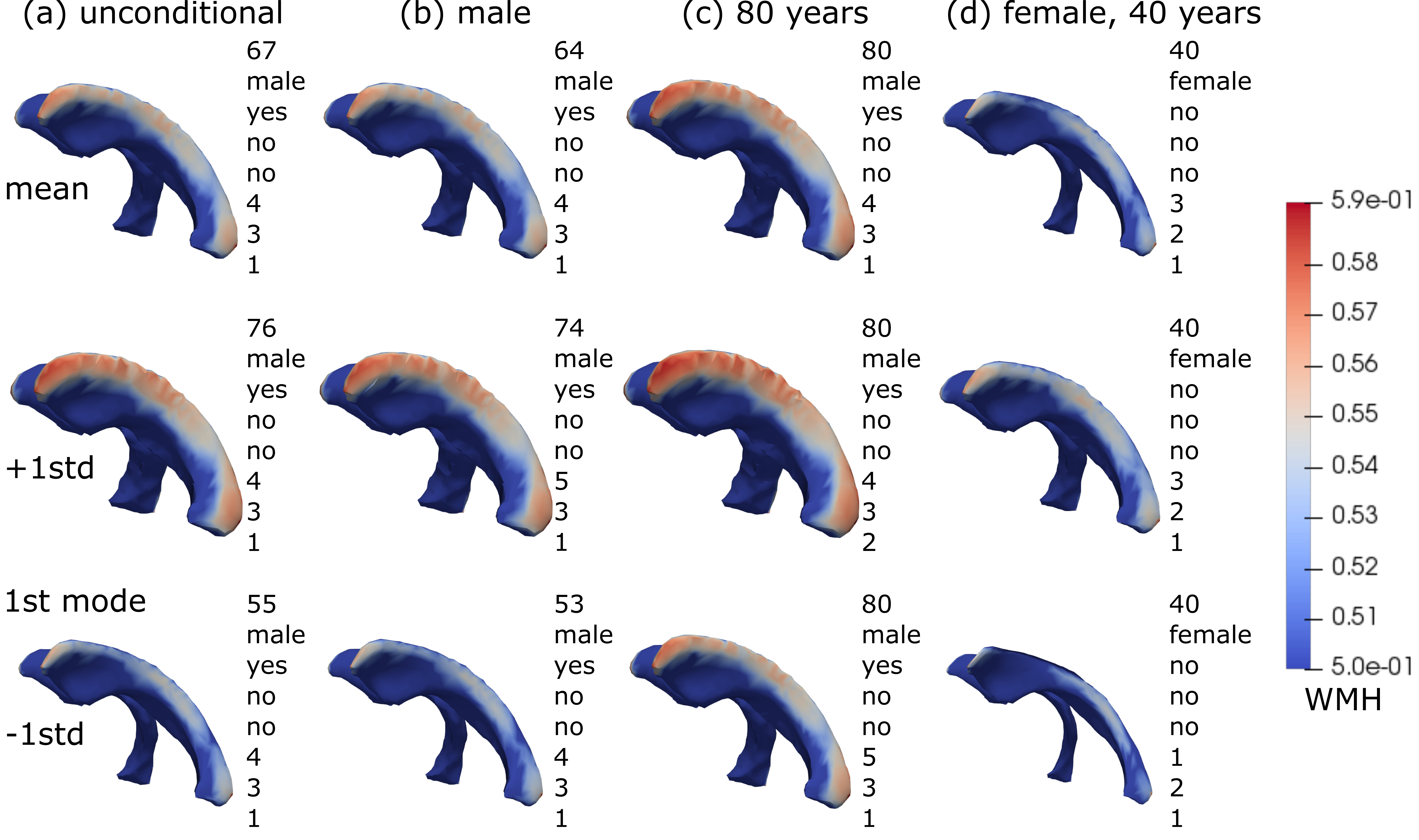}
\end{center}
\caption{Joint model of ventricle shape, WMH burden, and clinical indicators. Mean and variation along the first principal mode is shown. a) the original model, b), c), d) the model conditioned on $\rm{sex} = \rm{male}$, $\rm{age} = 80\: \rm{ years}$ and $(\rm{sex} = \rm{female}, \rm{age} = 40\:\rm{years})$ respectively. The color indicates the spatially varying amount of WMH burden. The variables per instance are ordered as age, sex, hypertension, hyperlipidemia, atrial fibrillation, smoking, NIHSS, and mRS.} 
\label{fig:variation}
\end{figure}

We produce a statistical model based on a dataset of T2-FLAIR magnetic resonance images from 793 acute ischemic stroke patients, without gross pathologies, such as mass effects, from the Genes Affecting Stroke Risk and Outcomes Study (GASROS), for which patients were enrolled at Massachusetts General Hospital between 2003 and 2011 \cite{zhang2015determinants}. Manual WMH segmentation was performed using MRIcro \cite{rost2010white}, and ventricles were automatically segmented (Dice: 0.89), using a 3D U-Net \cite{2019arXiv190700695D}. Subsequently, we isolated periventricular from subcortical WMH, based on connectivity of individual WMH lesions to the ventricles.  After segmentation, we establish correspondence between every ventricle segmentation and the atlas ventricles by fitting a Gaussian process deformation model estimated via the iterative closest point algorithm \cite{luthi2018gaussian}. The ventricles are represented as a surface triangle mesh with 1504 vertices. The WMH is modeled as a feature based on the vertices of the mesh. Every voxel labeled as periventricular WMH is assigned to the closest vertex point. This results in a WMH representation as a surface feature. In addition to the shape of the ventricles and the WMH burden feature, we include clinical indicators that represent age, sex, hypertension (binary), hyperlipidemia (binary), atrial fibrillation (binary),  smoking (ordinal, 1-6) and NIHSS (stroke score, ordinal 0-42), mRS (outcome score, ordinal 0-6) and the overall WMH volume. With 1504 3D ventricle vertices, 1504 WMH burden features and 9 clinical indicators, the total dimensionality of a single data point is $d=6025$.

Fig.~\ref{fig:variation} presents the joint model of the ventricle shape, WMH burden features, and clinical indicators and also three conditional models for different clinical indicators, like age and sex. The conditional models have similar mean and variability as the unconditioned model when conditioned on features with low correlations to the remaining components Fig.~\ref{fig:variation}b and show different means and less variability when conditioned on stronger correlated Fig.~\ref{fig:variation}c or combined features Fig.~\ref{fig:variation}d. The supplementary material contains a video\footnote{\href{https://youtu.be/gPoHP_iFQIA}{https://youtu.be/gPoHP\_iFQIA}} of our graphical user interface for data exploration. It enables the user to fix clinical indicators and visualize the remaining variability in the conditional model.
\begin{table}[t!]
\scriptsize
   \centering
  \caption{Clinical indicator prediction. Mean error and standard deviation are reported for the prediction of age, stroke outcome (mRS) and sex. The columns of the matrix correspond to the prediction based on the mean of the marginal distribution and conditioned on the ventricle shape, WMH burden, all other clinical indicators, indicators combined with WMH volume (ind + vol), and all available components except for the one being predicted. For sex, the percent of correct predictions is reported.}
 \label{tab:reconstruction}
 \begin{tabular}{|l|l|l|l|l|l|l|}
 \hline
 & mean & ventricles & WMH & indicators & ind + vol & combined\\
 \hline
Age & 15.77  $\pm$ 9.49& 11.27 $\pm$  6.62& 12.06 $\pm$  7.58& 13.21 $\pm$ 8.19 & 11.57 $\pm$  7.04 & \textbf{9.85} $\pm$ 5.76 \\
mRS & 2.01  $\pm$ 1.50 & 1.88 $\pm$ 1.37 & 1.96 $\pm$ 1.45 &  \textbf{1.65} $\pm$ 1.20 &  \textbf{1.65} $\pm$ 1.20 &  \textbf{1.65} $\pm$ 1.19 \\
Sex & 60.6\% & 62.7\% & 61.7\% & 65.3\%& 64.8\% & \textbf{67.0\%} \\

 \hline

 \end{tabular}
 \end{table}
  \begin{table}[t!]
\scriptsize
   \centering

   \caption{Ventricle Shape and WMH burden prediction. The setting is the same as in Table~\ref{tab:reconstruction} but we predict the shape of the ventricles and the WMH burden from the remaining values. Mean and standard deviation of distances between mesh vertices in mm is reported for the ventricle shape and WMH feature distance in voxels.}
    \label{tab:reconstructionShape}
 \begin{tabular}{|l|l|l|l|l|l|l|l|}
 \hline
 & mean &  ventricles & WMH & indicators & ind + vol & combined\\
 \hline
ventricles& 1582.2 $\pm$ 34.5 & - &     132.9  $\pm$ 35.0 & 146.9  $\pm$ 45.1& 144.8  $\pm$ 43.0&  \textbf{132.3} $\pm$ 34.8 \\
WMH & 3980.1 $\pm$ 2830 & 2927  $\pm$ 2270 & - &     3073 $\pm$ 2419 & 1291 $\pm$ 897 & \textbf{1254} $\pm$ 866 \\ 
 \hline

 \end{tabular}
 \end{table}

We evaluate the joint model in a reconstruction task of estimating the full vector from partial data in three different tasks: attribute prediction (Table~\ref{tab:reconstruction}), shape prediction, and WMH burden prediction (Table~\ref{tab:reconstructionShape}). For all tasks, we use the mean of the conditional distribution in the latent space to generate predictions. The values are obtained by randomly splitting the dataset into a training set of 600 patients and a validation set of the remaining 193 patients. The attribute prediction task (Table \ref{tab:reconstruction}) aims to estimate clinical indicators, e.g., stroke outcome, given the shape and WMH burden. We observe that the ventricle shape and WMH burden are strong indicators for age whilst they contribute little to the stroke outcome score. For age and sex, using the combination of all modalities provides the best prediction. The shape and WMH prediction tasks (Table \ref{tab:reconstructionShape}) reveal how the imaging data from a patient relates to the clinical indicators. The shape of the ventricles can be predicted from the WMH burden. The prediction in the other direction is more challenging, due to strong variability in WMH burden in the population. Clinical indicators including WMH volume combined with the ventricle shape produce the best prediction.

\begin{figure}[t!]
\begin{center}
\includegraphics[width=\textwidth]{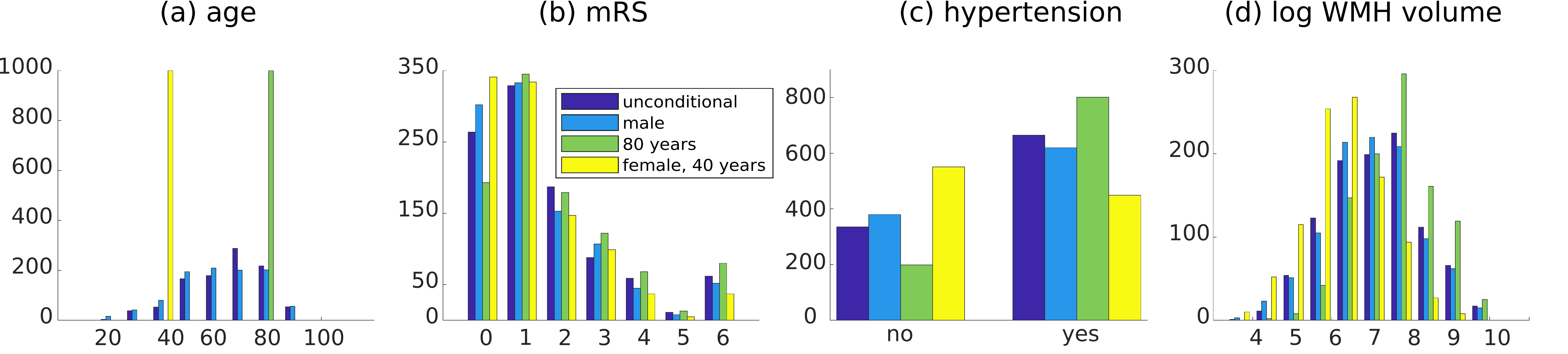}
\end{center}
\caption{Sample distributions  of age, mRS (outcome score), hypertension and the log WMH volume for unconditional and conditional model conditioned on $\rm{sex} = \rm{male}$, $\rm{age} = 80\: \rm{ years}$ and $(\rm{sex} = \rm{female, age} = 40\:\rm{years})$ respectively.} 
\label{fig:marginals}
\end{figure}

In the third experiment, we demonstrate the difference between the unconditional and conditional models. We produce the sample distributions by drawing 1,000 random samples from the corresponding model. Fig.~\ref{fig:marginals} reports those sample distributions before and after conditioning on the same three events as in Fig.~\ref{fig:variation}. We observe that the conditioning on male has a weak effect whilst the conditioning on age or a combination of age and sex has a strong effect on the resulting marginal distributions.

We implemented the copula model based on the code provided in \cite{egger17copula} in the Scalismo\footnote{Scalismo - A Scalable Image Analysis and Shape Modeling Software Framework \\
\href{https://github.com/unibas-gravis/scalismo}{https://github.com/unibas-gravis/scalismo}} framework for statistical shape modeling based on Gaussian processes. The computation of the conditional shape models based on the clinical indicators takes a few seconds for our dataset on a single core. Reducing the model via PCA will speed it up to real-time performance, enabling interactions with the model.

\textbf{Limitations} The proposed model has two main limitations. First, we assume dense correspondence on the surface of the ventricles as well as correspondence across image features. This is especially limiting if some shape features have weak or no correspondence across the population. Second, the Gaussian model for the dependency structure we are using is limited to second-order dependency and does not reflect higher-order dependency. These two limitations and assumptions are very common in the statistical shape modeling, especially when working with small datasets.

%A Third general limitation is model evaluation which is a challenging task itself. Egger \textit{et al.}~\cite{egger17copula} evaluated the shape and appearance part of a face model regarding specificity and generalization. Individual parts of the model can be evaluated that way, but those model metrics cannot be directly used to measure the joint model.

\section{Conclusion}
We present a joint generative model of shape, image features, and clinical indicators. Conditional models for partially observed data can be straightforwardly derived. The copula can be implemented as a simple pre- and postprocessing step in existing pipelines relying on (probabilistic) PCA or Gaussian processes. In addition, conditional models can be constructed efficiently for interactive data exploration. Our work provides a highly effective approach to exploring high dimensional data, especially spatial correlations, for example between ventricle expansion and WMH burden patterns. Statistical shape models are commonly used as statistical priors for image analysis tasks, and the proposed conditional models can be interpreted as patient-specific statistical shape models for patient-specific medical image analysis. Our model includes binary, discrete, ordinal and continuous variables and hence can handle a vast variety of available clinical indicators.

\textbf{Aknowledgments} This research was funded by SNSF P2BSP2\_178643, NIH NIBIB NAC P41EB015902, NIH NINDS R01NS086905, Horizon2020 753896, De Drie Lichten 24/18, ZonMw 104003005,  Wistron Corporation,
AWS, SIP and NVIDIA.

\bibliographystyle{splncs04}
\bibliography{paper1371}
\end{document}